\documentclass[aps,prd,nofootinbib]{revtex4}
\usepackage[colorlinks=true, pdfstartview=FitV, linkcolor=blue, citecolor=red, urlcolor=magenta]{hyperref}
\usepackage{graphicx}
\usepackage{latexsym}
\usepackage{amsmath}
\usepackage{amsfonts}
\usepackage{amssymb}
\usepackage{verbatim}
\usepackage{comment} 

\newcommand{\be}{\begin{equation}}
\newcommand{\ee}{\end{equation}}
\newcommand{\bea}{\begin{eqnarray}}
\newcommand{\eea}{\end{eqnarray}}

\newcommand{\pa}{\partial}

\newcommand{\ben}{\begin{eqnarray}}
\newcommand{\een}{\end{eqnarray}}

\begin{document}

\title{The 2D Lorentz-violating fermionic Casimir effect under thermal conditions}


\author{$^{1}$K. E. L. de Farias}
\email{klecio.lima@uaf.ufcg.edu.br}

\author{$^{1,3}$M. A. Anacleto}
\email{anacleto@df.ufcg.edu.br}

\author{ $^{2}$Iver Brevik}
\email{iver.h.brevik@ntnu.no}

\author{$^{1,3}$F. A. Brito}
\email{fabrito@df.ufcg.edu.br}

\author{$^{1,3}$E. Passos}
\email{passos@df.ufcg.edu.br}

\author{$^{1}$Amilcar Queiroz}
\email{amilcarq@df.ufcg.edu.br}

\author{$^{1,3}$João R. L. Santos}
\email{joaorafael@df.ufcg.edu.br}


\affiliation{$^{1}$Departamento de F\'{\i}sica, Universidade Federal de Campina Grande,\\
Caixa Postal 10071, 58429-900, Campina Grande, Para\'{\i}ba, Brazil.}
\affiliation{$^{2}$Department of Energy and Process Engineering, Norwegian University of Science and Technology\\  N-7491 Trondheim, Norway.}
\affiliation{$^{3}$Unidade Acad\^emica de Matem\'atica, Universidade Federal de Campina Grande,\\ 58429-970,  Campina Grande, Para\'{\i}ba, Brazil.}


\begin{abstract}
In the present work, we study a fermionic Lorentz invariance violation (LIV) theory with a CPT-even extension and analyze its impact on the Casimir effect under the MIT bag boundary condition model in a low-dimensional setting, where results are obtained without any approximations for a null-temperature system. Moreover, the Matsubara formalism is applied to derive closed expressions for the influence of temperature on the physical observables: Casimir energy, Casimir force, and entropy associated with the system in a LIV context. For each thermal observable, the influence of the LIV correction term is considered in the analysis of both low- and high-temperature regimes. Additionally, we construct a condensed matter analogue using the SSH model, where nonlinear fermionic dispersion and boundary-induced vacuum energy emerge, reproducing the analytical structure of the LIV Casimir effect.

\end{abstract}
 \maketitle


\section{Introduction}
Namely, the Casimir effect, first discovered by H. Casimir in 1948 \cite{Casimir:1948dh} and initially detected by Sparnaay in 1958 \cite{Sparnaay:1958wg}, essentially corresponds to limits on the frequencies of vacuum fluctuations between two plates, resulting in a difference in force between the interior and exterior of the plates. Over time, Casimir's prediction has been experimentally validated \cite{Sparnaay:1958wg,Lamoreaux:1996wh,Mohideen:1998iz,Bressi:2002fr} and has since been extended to various fields, underscoring its fundamental significance in quantum field theory \cite{milton2001}, condensed matter physics \cite{bordag2009}, cosmology \cite{Wang:2016och}, biophysics \cite{Machta2012, Pawlowski2013}, and even unconventional physics, such as Lorentz invariance violation (LIV) \cite{Cruz2017}. The main purpose of the present work is to discuss the influence of LIV on the Casimir effect.

 Many interesting investigations of the Casimir effect in the LIV context have appeared recently in the literature (see first Ref.\cite{Cruz2017}). The main point is that the parameters controlling LIV introduce corrections to the main quantities associated with the Casimir effect (namely, modifications to the energy spectrum and corrections to vacuum forces). For instance, several issues were addressed, such as {\it aether and aether-like Lorentz-violating scenarios} \cite{Chatrabhuti:2009ew,Santos:2017yov,Erdas_2021a,deFarias:2023xjf,Droguett:2024tpe}, {\it Horava\textendash{}Lifshitz-like theories} \cite{MoralesUlion:2015tve}, {\it thermal effects approach} \cite{Erdas_2021} and {\it Lorentz-violating with higher order derivatives} \cite{ADantas:2023wxd, deFarias:2024amf}. In our investigation, we shall consider a theory with high-derivative Lorentz-violating operators (which falls on an energy scale where LIV is more sensitive). 

For convenience, we consider a LIV higher-derivative operator capable of producing corrections with even exponents in the momentum for the associated fermionic dispersion relation. For this purpose, we follow the same procedure as Myers and Pospelov \cite{Myers:2003fd} and implement such a modification in the form: $\sim \psi (u\cdot\partial)^{3}u_{\mu}\gamma^{\mu}\bar\psi$ in the usual Dirac theory (being $u_{\mu}$ a constant-vector and $\gamma^{\mu}$ the two-dimensional gamma matrices). However, Casimir-LIV proposals involving higher-order derivative operators often result in expressions without exact analytic solutions, requiring approximations to compute the Casimir energy and force  (see, for example, Refs. \cite{MoralesUlion:2015tve,ADantas:2023wxd}). In order to deal with this question (generate exact expressions for energy and force), it was pointed out by Farias {\it et al.} \cite{deFarias:2024amf}, a two-dimensional higher derivative approach. Here, we revisited this investigation for a two-dimensional LIV-fermionic analogue theory. Namely, the significance of obtaining a closed-form solution lies in the ability to fully understand the behaviour of the system and explore its limits, thereby elucidating the physical constraints of the parameters involved.

The mathematical interpretation of the Casimir effect is given by introducing boundary conditions on the system, where they simulate the plates according to necessity. Usually, the Dirichlet and Neumann boundary conditions are used to explore the phenomenology of the Casimir effect. However, these boundary conditions present problems in describing effects in a fermionic context. Since fermionic fields obey the Dirac equation, their boundary conditions should be consistent with the behaviour of spinors. Therefore, the most adequate boundary condition in this approach is the MIT bag model, which is often used to confine fermions in a region \cite{Chodos:1974je,Donaire:2019nzy}. In addition to the boundary condition, temperature affects the system. Since there is no null temperature in the universe, the introduction of temperature plays an important role in understanding how quantum fluctuations are modified when they are increased. Among several methods to introduce the temperature as the thermal Hadamard function \cite{deFarias:2022rju} and the Thermal Field dynamics \cite{Leineker:2010uw, Costa:2010fx}, the Matsubara method \cite{matsubara1955new} was chosen due to implementation requirements of only the dispersion relation in a thermal equilibrium system as considered in the current work. 

Lattice models offer a powerful framework for simulating high-energy physics phenomena in condensed matter systems. Among them, the Su–Schrieffer–Heeger (SSH) model, Ref. \cite{su1979solitons}, stands out as a prototypical one-dimensional system that captures the essential features of Dirac fermions with sublattice symmetry. In this work, we investigate how the SSH model in the strong dimerisation regime (see Ref. \cite{asboth2016short}) gives rise to a nonlinear fermionic dispersion relation structurally similar to those found in Lorentz-violating (LIV) field theories. By imposing open boundary conditions (OBCs) on a finite SSH chain, we discretize the allowed momentum modes and evaluate the resulting Casimir energy. The OBCs used in our lattice model result in a discrete momentum spectrum characterized by half-integer quantization, analogous to that found in MIT bag boundary conditions for fermions in quantum field theory. While the physical interpretations differ, the mathematical similarity allows us to draw a connection between Casimir-like effects in condensed matter systems and those in confined relativistic field theories. 
This setup provides a natural condensed matter analogue for studying finite-size quantum vacuum effects in systems with modified dispersion, bridging lattice physics and LIV-inspired quantum field theory.

We present this paper in five sections, which are briefly described as follows.  In Section \ref{sec02}, we develop the fermionic sector with an extended higher derivative operator. Furthermore, we find the modified dispersion relation associated with the massless case, where we use natural units for simplicity in this section. 
In Section \ref{sec03}, we calculate the Casimir effect and discuss the influence of the LIV parameter on the Casimir energy and force, adopting SI units in this section. In the following section, \ref{sec04}, the influence of the temperature in the system is discussed, where the Casimir energy and the force dependent on the temperature are calculated, where expressions for low- and high-temperature limits are provided. The entropy associated with the system is calculated in Sec. \ref{sec05} considering the temperature limit cases presented in the Casimir energy and force. In Sec. \ref{sec06}, we study a condensed matter analogue based on the SSH lattice model, where a nonlinear fermionic dispersion relation emerges in the strong dimerization regime. By imposing boundary conditions on a finite chain, we compute the corresponding Casimir energy and show that it reproduces the analytical structure found in the LIV-model. Finally, in Sec. \ref{sec07}, we present our final comments and results.

\section{The 2D LIV-Fermion theories}\label{sec02}
Let us now introduce a 2D theory of the Dirac sector extended with the CPT-even LIV term. According to Ref. \cite{Liberati:2009pf}, the associated fermionic effective action is given by 
\bea\label{f01}
S^{2D}_{eff}= \int d^2x\,[i\bar\psi\pa_{\mu}\gamma^{\mu}\psi-i\bar{\alpha} \bar\psi(u\cdot \pa)^{3}u_{\mu}\gamma^{\mu}\psi]\,,
\label{d1}
\eea
where $\bar{\alpha} = \alpha/M^{2}_{\rm Pl}$, with $M^{2}_{\rm Pl}$ being the Planck mass of quantum gravity and $\alpha$ is a dimensionless LIV parameter responsible for introducing the high-derivative term. When the LIV parameter $\alpha$ is submitted to the limit $\alpha\to0$, any LIV effect vanishes and returns to the usual theory. Note that in the high-derivative order constructions, there are other contributions, namely $\sim1/M_{\rm pl}\bar\psi(u\cdot\pa)^{2}u_{\mu}\gamma^{\mu}\psi$, which corresponds to the lowest-order nonrenormalizable LIV CPT-odd operator \cite{Myers:2003fd}· However, this extension introduces a correction to the particle dispersion relation that vanishes in the context of the Casimir effect in 2D, then, its use is not applicable. Another CPT-even contribution that could be considered is given as $\sim1/M^2_{\rm pl}\bar\psi(u\cdot\pa)\partial^2u_{\mu}\gamma^{\mu}\psi$. However, this term can be reduced to lower-dimensional operators via the equations of motion and does not satisfy the necessary criteria for constructing a valid CPT-even higher-derivative operator (see \cite{Myers:2003fd}); therefore, it should be discarded.

Taking into account that the gamma matrices in 2D are defined as:
\bea\label{f06}
\gamma^{0}=  
\begin{pmatrix}
0\,& 1\\
1 \,& 0
\end{pmatrix},\;\gamma^{1}=  
\begin{pmatrix}
0\,& -1\\
1 \,&\;\; 0
\end{pmatrix}\,.
\eea
Therefore, the equation of motion associated to the 2D-effective action, Eq.(\ref{f01}), is rewritten in the following matrix representation
\bea\label{f04}
\begin{pmatrix}
0 & i\pa_{t}-i\pa_{x} - i\bar{\alpha}(u\cdot\pa)^{3}u_t+i\bar{\alpha}(u\cdot\pa)^{3}u_x\\
i\pa_{t}+i\pa_{x} - i\bar{\alpha}(u\cdot\pa)^{3}u_t-i\bar{\alpha}(u\cdot\pa)^{3}u_x& 0
\end{pmatrix}
\psi=0.
\eea


Note that from Eq. \eqref{f04}, we can derive a set of dispersion relations associated with the time-like and space-like configurations. In the momentum space $(\partial_t\to-i\omega,\partial_1\to ik_x)$, they are given, respectively, by
\begin{subequations}
    \bea\label{f12}
    \omega^{2} - k_{x}^{2} + 2\bar{\alpha} \omega^{4} +\bar\alpha^2\omega^6=0,\;\;\Leftrightarrow\;\;u_{t} = 1, u_x=0\,,
    \eea
    \bea\label{f13}
    \omega^{2} - k_{x}^{2} + 2\bar{\alpha} k_{x}^{4} - \bar{\alpha}^{2} k_{x}^{6} =0.\;\;\Leftrightarrow\;\;u_{t} = 0,u_x=1\,.
    \eea
\end{subequations}
Considering first the time-like case. Note that the dispersion relation associated with this configuration is not trivial, and gives us a set of solutions:
\begin{align}
   \omega =&-\left(\frac{2}{3}\right)^{1 / 3}\frac{1}{\left(9 k_x \bar{\alpha}^2+\sqrt{3} \sqrt{4 \bar{\alpha}^3+27 k_x^2 \bar{\alpha}^4}\right)^{1 / 3}}+\frac{\left(9 k_x \bar{\alpha}^2+\sqrt{3} \sqrt{4 \bar{\alpha}^3+27 k_x^2 \bar{\alpha}^4}\right)^{1 / 3}}{2^{1 / 3} 3^{2 / 3} \bar{\alpha}}\,,\label{dr1}\\
  \omega =&\frac{1\pm i \sqrt{3}}{2^{2 / 3} 3^{1 / 3}\left(9 k_x \bar{\alpha}^2+\sqrt{3} \sqrt{4 \bar{\alpha}^3+27 k_x^2 \bar{\alpha}^4}\right)^{1 / 3}}-\frac{(1\mp i \sqrt{3})\left(9 k_x \bar{\alpha}^2+\sqrt{3} \sqrt{4 \bar{\alpha}^3+27 k_x^2 \bar{\alpha}^4}\right)^{1 / 3}}{12^{2 / 3} \bar{\alpha}}\,.
  \label{dr2}
\end{align}
The previous expressions unveil that only Eq. \eqref{dr1} is a real solution. However, even this real solution presents some problems since it diverges as the LIV parameter $\alpha\to 0$, and we do not recover the usual result as expected. Since the parameter $\bar\alpha$ is small, as expected for the LIV effect, we introduce the following redefinition.
\begin{equation}
    y=\left(9 k_x \bar\alpha^2+\sqrt{3} \sqrt{4 \bar\alpha^3+27 k_x^2 \bar\alpha^4}\right)^{1 / 3}\,.
    \label{eqy}
\end{equation}
Hence, from the behaviour of the dispersion relation, Eq. \eqref{f12}, for $\omega$ when the coupling parameter is small, we can expand Eq. \eqref{eqy} in the smallness parameter $\bar\alpha$ up to next-to-leading order, one gets
\begin{equation}
y= \sqrt{\bar\alpha}(2\sqrt{3})^{1/3}
 \left( 1+\frac{3k_x}{2\sqrt{3}}\sqrt{\bar\alpha} \right).
 \label{eqy1}
\end{equation}
In this case, according to the \eqref{eqy1}, we can rewrite the expression in Eq. \eqref{dr1} as
\begin{equation}
\omega = \frac{1}{\sqrt{3\bar\alpha}}
\left[ \left( -1+ \frac{\sqrt{3}k_x}{2}\sqrt{\bar\alpha}\right) + \left( 1+ \frac{\sqrt{3}k_x}{2}\sqrt{\bar\alpha}  \right)     \right]\,,
\end{equation}
such as
\begin{equation}
\omega (k_x)= |k_x|.
\end{equation}
Therefore, it is a behaviour that the LIV does not exert any influence on. From Eq. (\ref{f13}), we can verify that the spacelike solution requires that
\bea\label{f15}
\omega(k_{x})=\pm\sqrt{k_x^2\left(1 - \bar\alpha k_x^2\right)^2}.
\eea
Notice that the above solutions correctly reproduce the usual dispersion relations in the limit $\bar\alpha \to 0$, Eq.(\ref{f15}), reduces to the following compact equation: 
    \bea\label{f17}
    \omega(k_{x})=|k_x|\left|1 - \bar\alpha k_x^2\right|.\;\;\Leftrightarrow\;\;u_{\mu} = (0,\vec{1})
    \eea
Hence, unlike the timelike case (in the small parameter, $\bar{\alpha}$,  approach), the spacelike case is affected by the LIV parameter.

\section{The Casimir energy}
\label{sec03}

Vacuum energy fluctuations arise from the inherent quantum fluctuations of fields, which can be interpreted as the brief creation and annihilation of virtual particle pairs, leading to measurable effects in the energy density of the vacuum. This energy may be manifested as the Casimir energy, which can be obtained by the modified dispersion found earlier and submitted to a boundary condition. To determine this energy in the proposed system, we have two alternatives based on the dispersion relations previously obtained. The first is the real solution of the timelike dispersion relation shown in \eqref{dr1}, which is the real result from the dispersion relation. However, it seems impossible to work out for two reasons. The first problem is the complexity of the form found, and the second point is the problem of the negative dispersion relation, as shown before. Moreover, it presents a divergence in the limit $\bar{\alpha}\to0$. Another alternative is the space-like approach, see Eq. \eqref{f15}, which gives us a good behaviour function with non-negative values. We also consider from now on the SI units, leading us to the following dispersion relation:
\begin{equation}
    \omega(k_{x})=\pm c\sqrt{k_x^2\left(1 - \frac{\alpha\hslash^2}{M_{\rm Pl}^2c^2} k_x^2\right)^2}\,,
\end{equation}
where the LIV parameter becomes $\bar\alpha=\frac{\alpha\hslash^2}{M_{\rm Pl}^2c^2}$. By taking the MIT bag boundary condition in the $ z$-direction, the momentum component becomes discrete, with $k_n=\left(n+\frac{1}{2}\right)\frac{ \pi}{a}$, $ n=0, 1, 2, \ldots$ for a massless case. Consequently, the vacuum energy may be derived by using: 
\begin{align}
E_0(a)=-\hslash \sum_n\omega_n,
\label{c1}
\end{align}
which means the sum of all normal modes of a wave oscillation. The fluctuation energy under a MIT bag boundary condition is written as
\begin{align}
   E_0(a) &=-\hslash c\sum_{n=0}^{\infty}\sqrt{\left(n+\frac{1}{2}\right)^2\left(\frac{ \pi}{a}\right)^2\left[1-\bar{\alpha}\left(n+\frac{1}{2}\right)^2\left(\frac{ \pi}{a}\right)^2\right]^2}\nonumber\\
   &=  - \frac{\hslash c}{\bar{\alpha}^{1 / 2}}\sum_{n=0}^{\infty} \sqrt{\bar{g}^2 \left(n+\frac{1}{2}\right)^2\left(1-\bar{g}^2 \left(n+\frac{1}{2}\right)^2\right)^2} ,
   \label{c2}
\end{align}
where $\bar{g}^2= \bar{\alpha}\frac{\pi^2}{a^2}$ is a adimensional parameter.

We can verify that the sum present in the above expression is divergent. To avoid such a divergence, we have to use a regularization method to separate the divergent part from the finite energy term originating from the Casimir energy. Here, we use the Abel-Plana formula, but first, the previous sum should be rewritten as 
\begin{equation}
    f(n)=\sqrt{\bar{g}^2n^2\left[1-\bar{g}n^2\right]^2}\,.
    \label{c6}
\end{equation}
Some consideration should be taken before the calculus. Then, the Abel-Plana formula to an integer sum is given by
\begin{equation}
\sum_{k=0}^{\infty} f(k+1/2)=\int_0^{\infty} d x f(x)-i \int_0^{\infty} d x \frac{f(i x)-f(-i x)}{e^{2 \pi x}+1}\,.
\label{c7}
\end{equation}
The advantage of using this method is that the divergent part is presented in the first integral in the r.h.s., which is discarded by the renormalization process. Therefore, the normalized Casimir energy is found by performing the last integral of the r.h.s, which gives us the following result
\begin{align}
E_0^{\rm ren}(a)&=\frac{i\hslash c}{\bar{\alpha}^{1 / 2}} \int_0^{\infty} \frac{d x}{e^{2 \pi x}+1}\left\{\sqrt{\bar{g}^2(i x)^2\left[1-\bar{g}^2(i x)^2\right]^2}-\sqrt{\bar{g}^2(-i x)^2\left[1-\bar{g}^2(-i x)^2\right]^2}\right\}\nonumber \\
& =-\frac{2\hslash c}{\bar{\alpha}^{1 / 2}} \int_0^{\infty} \frac{d x}{e^{2 \pi x}+1} \sqrt{\bar{g}^{2} x^2\left[1+\bar{g}^{2} x^2\right]^2}=-\frac{\pi  \hbar c}{24 a}-\frac{7 \pi ^3 \bar{\alpha}  \hbar c}{960
   a^3}.
   \label{c8}
\end{align}
It is straightforward to note that in the limit $\bar{\alpha}\to0$, the LIV contribution vanishes, and the first term is the usual result of a unidimensional fermionic system. Notice also that the LIV contribution is proportional to $\mathcal{O}(3)$ in the distance, while the usual term has $\mathcal{O}(1)$. This result unveils that at a large distance, the usual parts of the Casimir energy dominate, while in the short distance, the LIV term dominates, allowing for finding a signal of the LIV presence in an experiment. Moreover, the energy diverges as $a\to0$ as expected. 

The presence of a boundary condition in the system leads us to a Casimir energy density, which is a finite energy present between the plates of the system. However, the energy cannot be measured in the laboratory. Nevertheless, this energy acts on the plates causing a force, which may be attractive or repulsive depending on the boundary condition to which the system is exposed. This force is found by taking the negative derivative of the energy in terms of the distance, i.e.:

\begin{equation}
    F_c=-\frac{d E}{da}=-\frac{\pi  \hslash c}{24 a^2}-\frac{7 \pi ^3 \bar{\alpha}  \hslash c}{320
   a^4}.
    \label{f1}
\end{equation}
Note that the force is negative, which leads us to an attractive force between the points. It is worth highlighting that both Casimir's energy density and Casimir's force present a linear behaviour in $\bar{\alpha}$, and the expression obtained does not need any approximation, which is usually necessary in theories with higher order derivative terms. Furthermore, the behaviour of the force is similar to the energy density, where the force dissipates when $a\to\infty$ and diverges to $a\to0$, but with greater intensity, once the order of the factors is increased in one. For both the energy and Casimir force cases, the LIV correction term contributes to the increase in intensity.


\section{Finite temperature}
\label{sec04}

In this section, we analyze the effect of the temperature on the system. There are many ways to introduce temperature into a model, and here, we used the Matsubara formalism. In this formalism, we should use a Euclidean field theory, which is obtained by a Wick rotation in the time coordinate, $t \rightarrow-i \tau$, consequently, the Euclidean time $\tau$ is confined to the interval $\tau \in[0, \beta]$, where $\beta=(k_B T)^{-1}$, being $k_B$ the Boltzmann constant. The partition function in the path integral representation becomes:
\begin{equation}
    Z=\quad \int \quad D \bar{\psi} D \psi \exp \left(-\int_0^\beta d \tau \int d^3 x \mathcal{L}_E\right) .
    \label{t1}
\end{equation}
As we consider a free fermion field, the partition function is simplified as
\begin{equation}
    Z=\operatorname{det}\left(\gamma_E^\mu \partial_{\mu E}+m\right)\,.
     \label{t2}
\end{equation}
Hence, by using the partition function above, we may find the free energy given by the following expression:
\begin{equation}
    \mathcal{F}=-k_BT \ln \left[\operatorname{det}\left(\gamma_E^\mu \partial_{\mu E}+m\right)\right]=- k_BT \operatorname{Tr}\left[\ln \left(P_E^2+m^2\right)\right],
     \label{t3}
\end{equation}
where $P_E^2$ is the Dirac operator.

The following expression gives the finite temperature Casimir effect:
\begin{equation}
E_{\rm cas}(a, T)=\mathcal{F}(a, T)-\mathcal{F}(T),
\label{t4}
\end{equation}
where $\mathcal{F}(a,T)$ is the free energy of the system, and 
\begin{equation}
\mathcal{F}(T)=E_0+f(T), 
\label{t5}
\end{equation}
the first term of the r.h.s. is the Minkowski divergent term at null temperature, and the second one is the radiation term. The free energy is given by
\begin{equation}
\mathcal{F}(T, a)=-\frac{1}{\beta} \sum_{l=-\infty}^{\infty\prime} \sum_{n=0}^{\infty} \ln \left(\xi_l^2+\omega_{n}^2\right) \text {, }
 \label{t6}
\end{equation}
where the prime sign means the odd values of $\ell$, i.e.,
\begin{equation}
\xi_l=\frac{\pi}{\beta}\ell, \quad \ell= \pm 1, \pm 3, \pm 5, \ldots\,.
 \label{t7}
\end{equation}
We can see that $\xi_{\ell}$ are the Matsubara frequencies for a fermion system. The discrete values come from the anti-periodic boundary condition applied to the $\tau$. Therefore, the free energy may be rewritten as
\begin{align}
 \mathcal{F}(T, a)&=-\frac{1}{\beta} \sum_{l=-\infty}^{\infty\prime} \sum_{n=0}^{\infty} \ln \left(\xi_l^2+\omega_n^2\right) =\frac{1}{\beta} \lim _{s \rightarrow 0} \frac{\partial}{\partial s} \frac{1}{\Gamma(s)} \sum_{l=-\infty}^{\infty\prime} \sum_{n=0}^{\infty} \int_0^{\infty} d t\ t^{s-1} e^{-t\left[\frac{\pi^2}{\beta^2}(2 l+1)^2+\omega_n^2\right]}.
 \label{t8}
\end{align}
Observe that the sum over $\ell$ is not trivial. Before any calculus, we should rearrange the sum by considering the recurrence relation: 
\begin{equation}
\sum_{l=-\infty}^{\infty\prime} e^{-t\left(\frac{\ell \pi}{\beta}\right)^2}=\frac{\beta}{2 \sqrt{\pi t}}+\frac{\beta}{\sqrt{\pi t}} \sum_{\ell=1}^{\infty}(-1)^{\ell} e^{-\frac{\ell^2 \beta^2}{4 t}}.
\label{t9}
\end{equation}
Consequently, Eq. \eqref{t8} may be rewritten in the following form:
\begin{align}
\mathcal{F}(T, a)=\lim _{s \rightarrow 0}\frac{\partial}{\partial s} \frac{1}{\Gamma(s)} \int_0^{\infty} d t \frac{t^{s-1}}{\sqrt{\pi t}}\left\{\frac{1}{2 }+ \frac{1}{2}\sum_{n=1}^{\infty} e^{-t \omega_n^2} +\sum_{\ell=1}^{\infty}(-1)^{\ell}e^{-\frac{\ell^2 \beta^2}{4 t}}+\sum_{n=1}^{\infty} \sum_{\ell=1}^{\infty}(-1)^{\ell} e^{-\frac{\ell^2 \beta^2}{4 t}-t \omega_n^2}\right\}
\label{t11}
\end{align}
The first term in parentheses represents the result that is obtained when we consider the Minkowski spacetime, which gives us a divergent result, which will be removed at the given moment. Note that we can compact it as a contribution $n=0$ in the sum present in the second term in parentheses, and then we obtain the same result present in Eq. \eqref{c8}. The next term (third in the above equation) is obtained when $n=0$ from the last terms in the bracket, leading us to the black body radiation term $f(T)$, which comes from the Minkowski thermal solution, and we remove it in a thermal renormalization process.

The last one is the contribution of the temperature and the boundary condition at the same time. It is expressed by

\begin{align}
     \lim _{s \rightarrow 0} \frac{1}{\Gamma(s)} \int_0^{\infty} d t \frac{t^{s-1}}{\sqrt{\pi t}} \sum_{n=1}^{\infty} \sum_{\ell=1}^{\infty}(-1)^{\ell} e^{-t\omega_n^2-\ell^2 \beta^2 / 4 t}
      &=\lim _{s \rightarrow 0} \frac{\partial}{\partial s} \sum_{n=1}^{\infty} \sum_{\ell=1}^{\infty}\frac{(-1)^{\ell}2^{\frac{3}{2}-s}}{\sqrt{\pi}}\left(\frac{\omega_n^2}{ \ell^2 \beta^2}\right)^{\frac{1}{4}-\frac{s}{2}} \frac{K_{\frac{1}{2} -s}\left(2 \ell\beta\sqrt{\omega_n^2}\right)}{ \Gamma(s)}\nonumber \\
      &=\frac{1}{2\beta} \sum_{n=1}^{\infty} \sum_{\ell=1}^{\infty}(-1)^{\ell} \frac{e^{-2 \ell \beta \omega_n}}{\ell}\,.
      \label{t14}
\end{align}

Finally, substituting the above results in Eq. \eqref{t4}, we obtained the renormalized thermal Casimir energy associated with the 
\begin{align}
E_{\rm cas}(T,a)=-\frac{\pi  \hslash c}{24 a}-\frac{7 \pi ^3 \bar{\alpha}  \hslash c}{960
   a^3} +\frac{1}{2\beta} \sum_{n=1}^{\infty} \sum_{\ell=1}^{\infty}(-1)^{\ell} \frac{e^{-2 \ell \beta \omega_n}}{\ell}.
\label{t15}
\end{align}
Note that the first terms are independent of the temperature parameter, then, in the analysis of temperature behaviour, these terms play the role of a constant, and they may be removed in the thermal analysis.

\subsection{Low temperature limit}

As mentioned before, the order in which the sums are performed shows us an expression that may be used to find the low-temperature behaviour, $T\to0(\beta\to\infty)$. Therefore, if the sum in $n$ is first performed, it does not give an analytic result. Then, we must consider some approximation since the LIV parameter $\bar{\alpha}$ is very small compared to the boundary condition distance $a$. It is worth highlighting that the null temperature term may be removed since it is a constant in this limit. Then by expanding Eq. \eqref{t15} in respect to $\bar{\alpha}$, we get

\begin{align}
E_{\rm cas}^{\rm low}(T,a)&=\frac{1}{2\beta} \sum_{n=1}^{\infty} \sum_{\ell=1}^{\infty} (-1)^{\ell}\frac{e^{-2 \ell \beta \omega_n}}{\ell} \nonumber\\
&\approx \sum_{n=1}^{\infty}\sum_{\ell=1}^{\infty}\frac{(-1)^{\ell}}{\beta}\left[\frac{\pi  e^{-2 \hslash c\beta  l \left(n+\frac{1}{2}\right)\frac{\pi}{a}}}{ l}+\frac{2 \pi ^3 \bar{\alpha}  \hslash c\beta  \left(n+\frac{1}{2}\right)^3 e^{-2 \hslash c\beta  l
   \left(n+\frac{1}{2}\right)\frac{\pi}{a}}}{a^3}\right].
   \label{l1}
\end{align}
Performing the sum in $n$, we obtain
\begin{align}
E_{\rm cas}^{\rm low}(T,a)=\sum_{\ell=1}^{\infty}\frac{\pi  (-1)^l e^{-l\pi\hslash c\beta/a } }{\beta}\left[\frac{1}{ l \left(e^{2 l\pi\hslash c\beta/a}-1\right)}+\frac{\pi ^2 \bar{\alpha}  \hslash c\beta   \left(5 e^{2 l\pi\hslash c\beta/a}+17 e^{4 l\pi\hslash c\beta/a}+27 e^{6 l\pi\hslash c\beta/a}-1\right)}{4 a^3 \left(e^{2 l\pi\hslash c\beta/a}-1\right)^4}\right].
\label{l2}
\end{align}

In the low-temperature limit, we consider $T\to0\ (\beta\to\infty)$, then, the above expression becomes:
\begin{align}
    E_{\rm cas}^{\rm low}(T,a)\approx \sum_{\ell=1}^{\infty}\frac{\pi  (-1)^l e^{-l\pi\hslash c\beta/a} }{\beta}\left[\frac{1}{l e^{2 l\pi\hslash c\beta/a}}+\frac{27\pi ^2 \bar{\alpha}  \hslash c\beta  e^{-2 l\pi\hslash c\beta/a}}{4 a^3}\right].
    \label{l3}
\end{align}

\begin{figure}[h]
    \centering
    \includegraphics[width=0.5\linewidth]{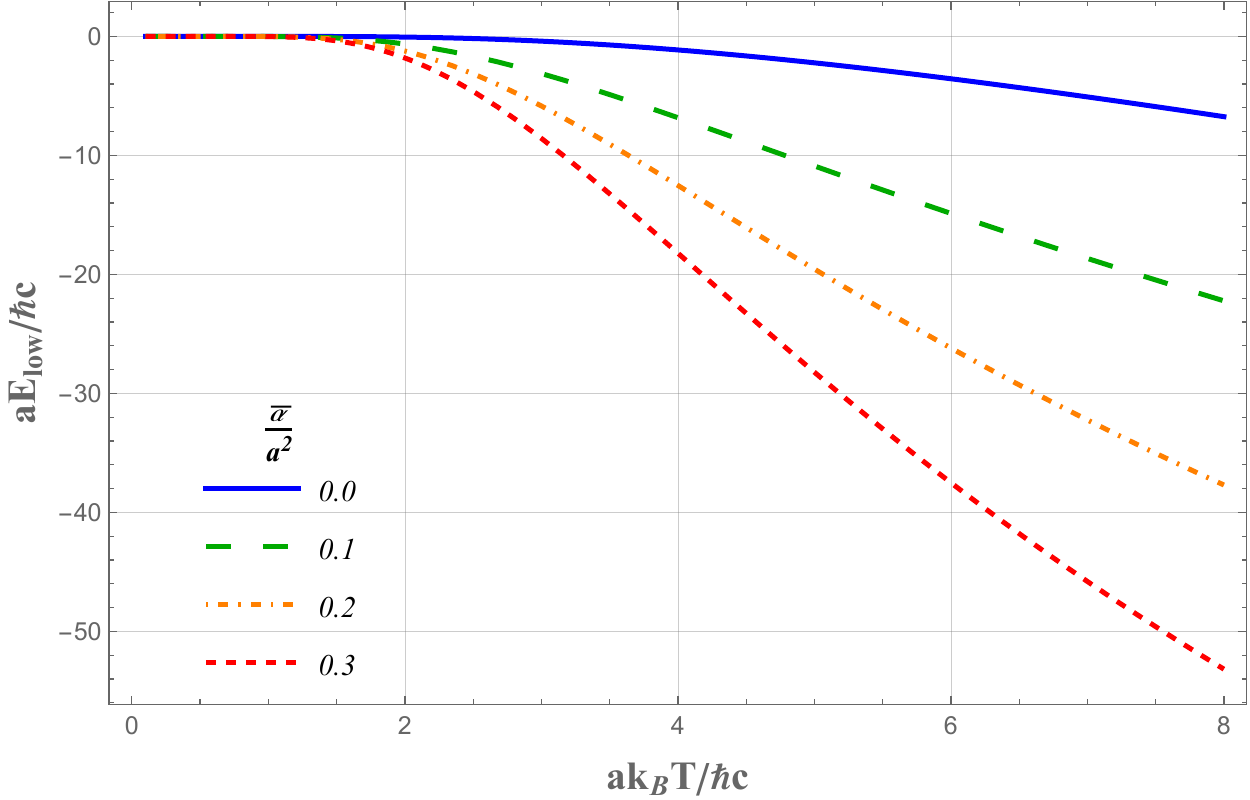}
    \caption{Plot of the dimensionless energy $aE_{\rm cas}^{\rm low}(T,a)/\hslash c$ in terms of $ak_{\rm B}T/\hslash c$ with a fixed distance $a$.}
    \label{eneL}
\end{figure}

Finally, the sum in $\ell$ may be performed and provides us with the following expression

\begin{align}
    E_{\rm cas}^{\rm low}(T,a) =-\frac{\pi  \log \left(e^{-3 \pi\hslash c\beta/a
   }+1\right)}{\beta}-\frac{27\hslash c\pi ^3 \bar{\alpha} }{4
   a^3 \left(e^{3\pi\hslash c\beta/a }+1\right)}.
   \label{l4}
\end{align}
By taking the limit $T\to0$, the thermal contribution vanishes as expected. This behaviour may be observed clearly in Fig. \ref{eneL}.

It is straightforward to note from Fig. \ref{eneL} that the Casimir energy at low temperatures is extremely sensitive to the increase in the LIV parameter $\alpha$. In addition, the behaviour of the Casimir energy with LIV presence and without LIV remains similar, but the LIV parameter modifies the intensities of each energy. It is worth highlighting that the energy from the vacuum fluctuation cannot increase infinitely as $T\to\infty$. This behaviour is a consequence of the approximation used in the low-temperature regime. Consequently, it should present an upper limit after a determined temperature.

\begin{figure}[h]
    \centering
    \includegraphics[width=0.5\linewidth]{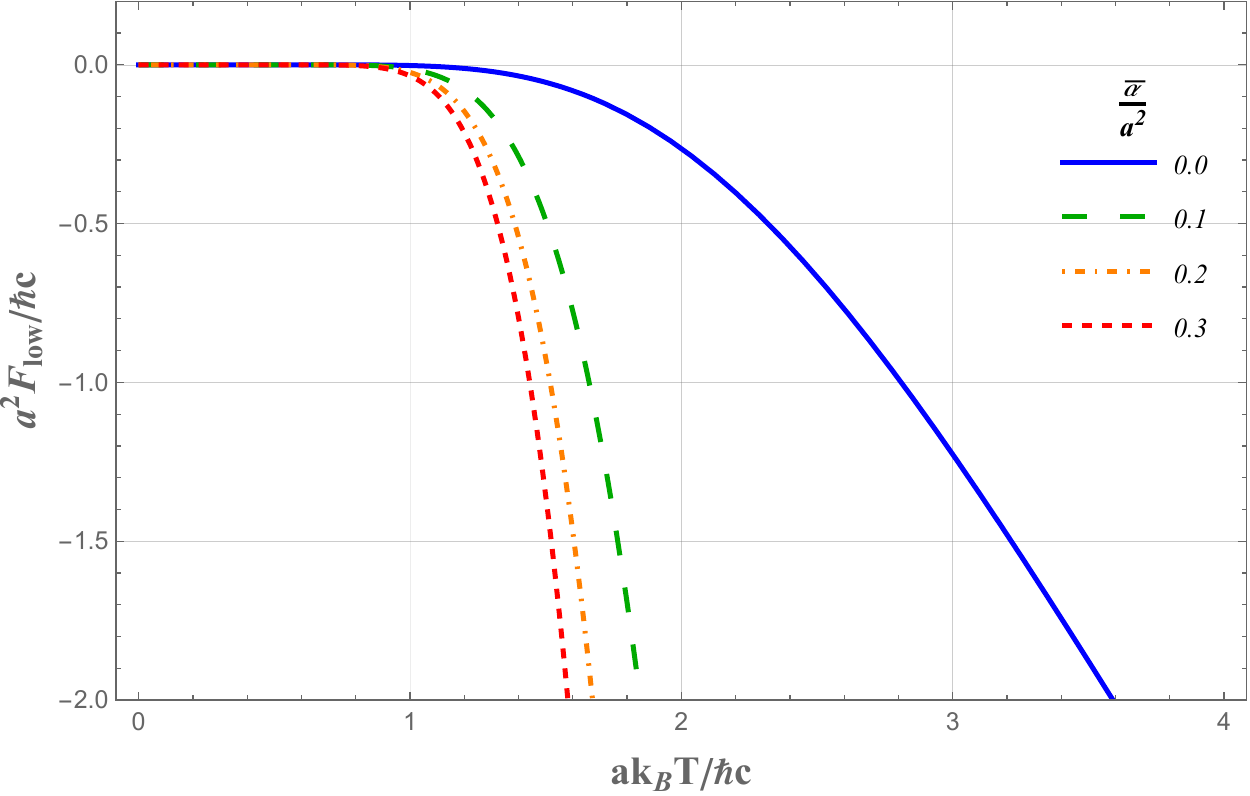}
    \caption{Plot of the dimensionless Force $a^2F_{\rm cas}^{\rm low}(T,a)/\hslash c$ in terms of $ak_{\rm B}T/\hslash c$ with a fixed distance $a$.}
    \label{forcL}
\end{figure}

The force may be obtained from the energy equation. The derivative in terms of $a$ from the expression without approximation gives us a long equation with many terms of each order in the exponential function. Therefore, we are going to focus on a simplified version of a low-temperature approximation, whose equation is
\begin{equation}
   F(T,a)=-\frac{3 \hbar c \pi^2}{a^2\left(1+e^{\frac{3 \hbar c \pi \beta}{a}}\right)^2}\left[1-\frac{27 \pi \bar\alpha}{4a^2}+e^{\frac{3 \hbar c \pi \beta}{a}}\left(1-\frac{27 \pi \bar\alpha}{4a^2}+\frac{27 \hbar c \pi^2 \bar\alpha \beta}{4a^3}\right)\right].
   \label{l5}
\end{equation}
As expected, in the limit $T\to0$, this expression vanishes completely. Moreover, the expression obtained is inadequate to obtain a high-temperature, as shown in Fig. \ref{forcL}.

It is worth noting that there is a point where the force in the presence of LIV $\bar{\alpha}\neq0$ and without LIV $\bar{\alpha}=0$ coincides regardless of the symmetry breaking value. The behavior of the curve shows that the force follows a pattern where it presents a minimum limit associated with each value of $\bar{\alpha}/a^2$. Furthermore, the increase in the LIV parameter causes the temperature required for a transition from attractive to repulsive behavior to be lower.


\subsection{High Temperature Limit}

The high-temperature limit is obtained by taking the sum over $\ell$ in a regime where $T\to\infty$, which corresponds to $\beta\to0$. By following the previous procedures, we should remove the null temperature term since it does not contribute to the thermal behaviour. Hence, from Eq. \eqref{t15}, the sum over $\ell$ results in
\begin{align}
E_{\rm cas}(T,a)=-\frac{1}{\beta} \sum_{n=1}^{\infty} \ln(1+e^{-2 \beta \omega_n})\,.
 \label{h1}
\end{align}
This result is well-known in the literature as the closed form of the fermion at finite temperature. However, the sum over $n$ of the above equation is divergent, consequently, we must use the approximation $\ln (1+z)= \approx z$ since $e^{-2 \beta \omega_n}<1$, then 
\begin{align}
E_{\rm cas}^{\rm high}(T,a)&=-\frac{1}{\beta} \sum_{n=1}^{\infty} \ln \left(1+e^{-2 \beta \omega_n}\right)\nonumber\\ &\approx-\frac{1}{\beta} \sum_{n=1}^{\infty} e^{-2 \beta \omega_n}\,.
\label{h2}
\end{align}
The renormalized part of the energy given by the above sum may be obtained by taking the Abel-Plana formula \eqref{c7}, consequently, we get 
\begin{equation}
E_{\rm ren}^{\rm high}(T,a)= \sum_{n=1}^{\infty}\frac{i}{\beta} \int_0^{\infty} \frac{d t}{e^{2 \pi t+1}}\left\{e^{-2 \beta \omega_{n+}}-e^{-2 \beta \omega_{n-}}\right\},
\label{h3}
\end{equation}
where
\begin{equation}
\omega_{i \pm }= \pm i \frac{t\hslash c\pi}{a}\left[1+\bar{\alpha} t^2\left(\frac{\pi}{a}\right)^2\right].
\label{h4}
\end{equation}
Therefore, the Casimir energy density takes the following form:
\begin{align}
E_{\rm ren}^{\rm high}(T,a)=& -\frac{i}{\beta} \int_0^{\infty} \frac{d t}{e^{2 \pi t}+1} 2 i \sin \left\{\frac{2\hslash c\beta t \pi}{a}\left[1+\bar{\alpha}t^2\left(\frac{\pi}{a}\right)^2\right]\right\}\nonumber \\
& =\frac{2}{\beta} \int_0^{\infty} \frac{d t}{e^{2 \pi t}+1}\left\{\sin \left(\frac{2\hslash c\beta t \pi}{a}\right) \cos \left[2 \hslash c\beta \bar{\alpha}\left(\frac{t \pi}{a}\right)^3\right]+\cos \left( \frac{2\hslash c\beta t \pi}{a}\right) \sin \left[2 \hslash c \beta \bar{\alpha}\left(\frac{t \pi}{a}\right)^3\right]\right\}.
\label{h5}
\end{align}
As the LIV constant $\bar{\alpha}$ is much smaller than the unit, i.e., $\bar{\alpha}\ll1$, we may consider an approximation and Eq. \eqref{h5} is rewritten as
\begin{equation}
E_{\rm ren}^{\rm high}(T,a)=\frac{2}{\beta} \int_0^{\infty} \frac{d t}{e^{2 \pi t}+1}\left\{\sin \left(\frac{2\hslash c\beta t \pi}{a}\right)+2 \hslash c\beta \bar{\alpha}\left(\frac{t \pi}{a}\right)^3 \cos \left(\frac{2 \hslash c\beta t \pi}{a}\right)\right\}.
\label{h6}
\end{equation}
Finally, by performing the integrals, we find an analytic expression which describes the renormalized Casimir energy in a high-temperature regime:
\begin{align}
E_{\rm cas}^{\rm high}(T,a)=\frac{a}{2 \pi \hslash c\beta ^2}+\frac{3 a \bar{\alpha} }{4 \pi (\hslash c)^3 \beta ^4}-\frac{\text{csch}\left(\frac{\pi\hslash  c\beta }{a}\right)}{2 \beta } -\frac{\pi ^3 \hslash c\bar{\alpha}  \left(23 \cosh \left(\frac{\pi\hslash c \beta }{a}\right)+\cosh \left(\frac{3 \pi \hslash c\beta
   }{a}\right)\right) \text{csch}^4\left(\frac{\pi\hslash  c\beta }{a}\right)}{32 a^3}\,.
\label{h7}
\end{align}
In a high-temperature regime, we should use the expansion to small arguments in the hyperbolic function, which gives us
\begin{equation}
   E_{\rm cas}^{\rm high}(T,a)=\frac{\pi \hslash c}{12 a} + \frac{7 \pi ^3 \hslash c\bar{\alpha} }{480 a^3}-\frac{(\hslash c)^3\beta ^2}{a^3} \left(\frac{7 \pi ^3}{720}+\frac{31 \pi ^5 \bar{\alpha} }{2016 a^2}\right)\,.
   \label{h8}
\end{equation}

\begin{figure}[h]
    \centering
    \includegraphics[width=0.5\linewidth]{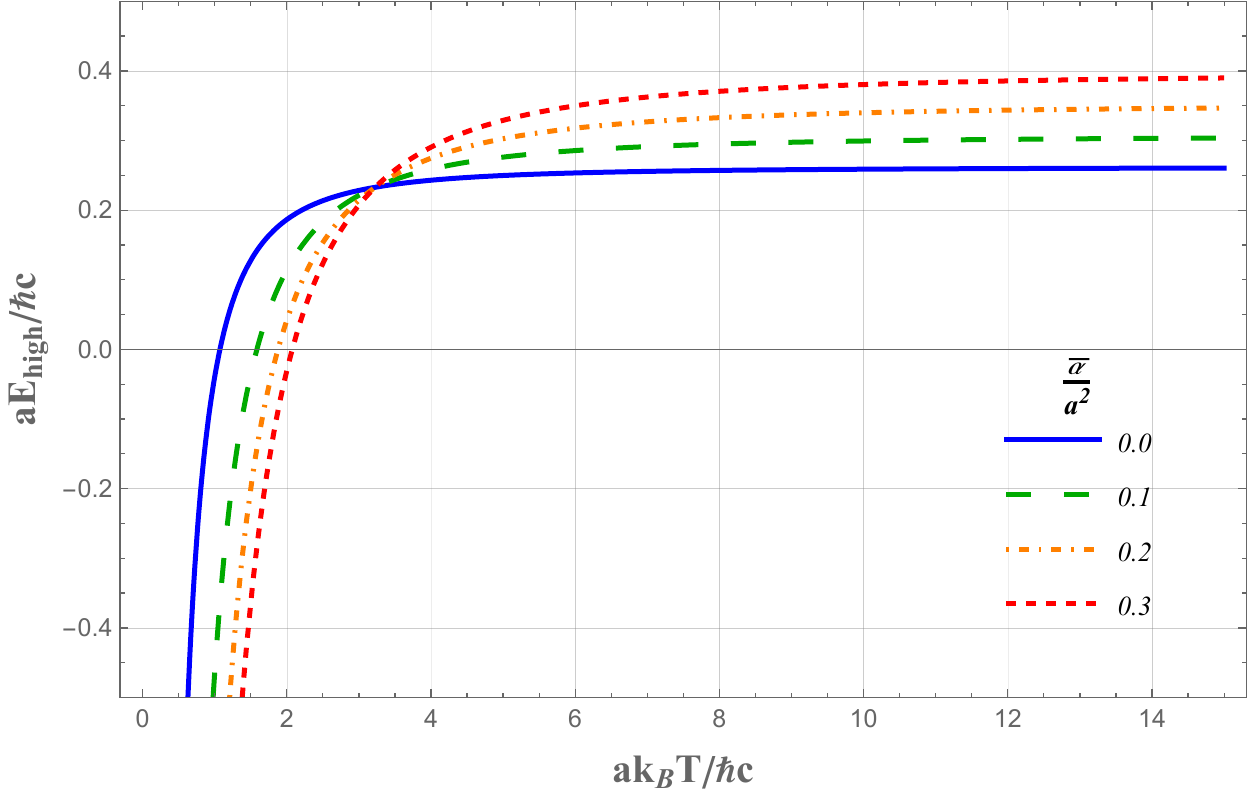}
    \caption{Plot of the dimensionless energy $aE_{\rm cas}^{\rm high}(T,a)/\hslash c$ in terms of $ak_{\rm B}T/\hslash c$ with a fixed distance $a$.}
    \label{enerH}
\end{figure}

\begin{figure}[h]
    \centering
    \includegraphics[width=0.5\linewidth]{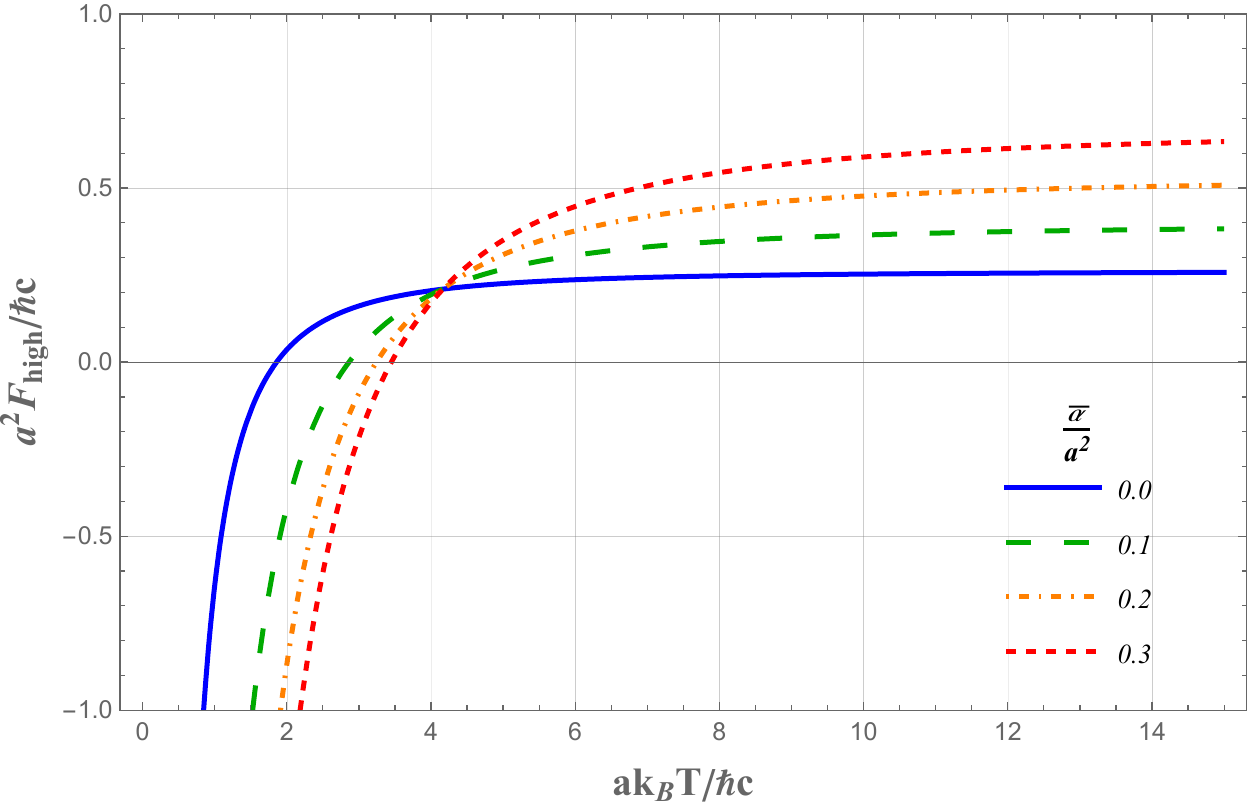}
    \caption{Plot of the dimensionless Force $a^2F_{\rm cas}^{\rm high}(T,a)/\hslash c$ in terms of $ak_{\rm B}T/\hslash c$ with a fixed distance $a$.}
    \label{forH}
\end{figure}

And the force will be expressed as

\begin{equation}
   F_{\rm cas}^{\rm high}(T,a)=\frac{\pi\hslash c}{12 a^2}+\frac{7 \pi ^3\hslash c\bar{\alpha} }{160 a^4}- \frac{(\hslash c)^3\beta
   ^2 }{a^4}\left(\frac{7 \pi ^3}{240}+\frac{155 \pi ^5 \bar{\alpha} }{2016
   a^2}\right)\,.
   \label{h9}
\end{equation}
Note that the behaviour of the Casimir energy and the Casimir force \eqref{h9} are similar in contrast to the expressions found at a low-temperature regime, consequently, both present a similar behaviour which diverges for $a\to0$ and vanishes for $a\to\infty$. Moreover, the LIV part of both energy and force is more relevant at short distances since it diverges faster than the non-LIV part. The complete behaviour of the Casimir force is shown in Fig. \ref{forH}. 

In Fig. \ref{forH}, it is possible to see that the force converges to a constant value after reaching a critical temperature. Such a behaviour may be explained by the fact that although the temperature can be as high as possible, the quantum fluctuations are limited and have a maximum value which becomes independent of any temperature. 


\section{Entropy}
\label{sec05}

\begin{figure}[h]
    \centering
    \includegraphics[width=0.5\linewidth]{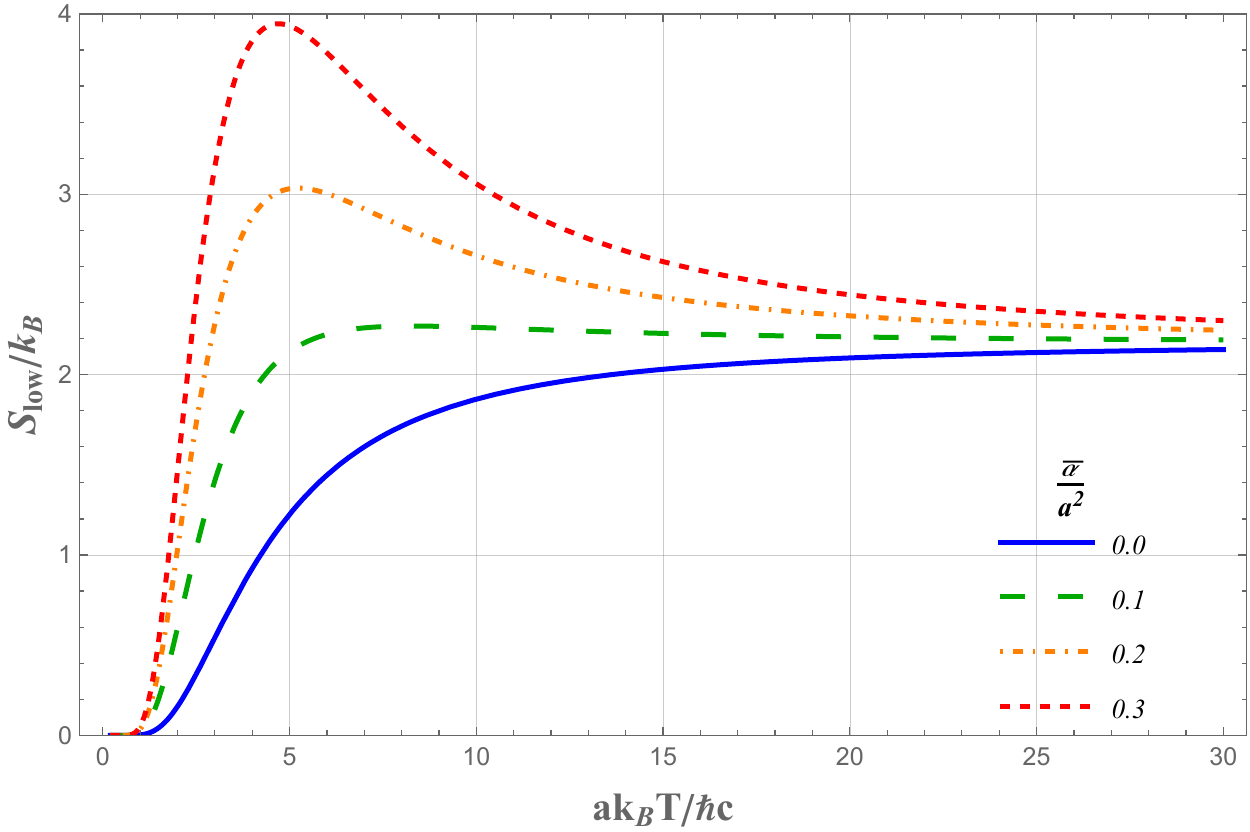}
    \caption{Plot of the dimensionless entropy $S_{\rm cas}^{\rm low}(T,a)/k_{\rm B} $ in terms of $ak_{\rm B}T/\hslash c$ with a fixed distance $a$.}
    \label{entroL}
\end{figure}

Since we have the thermal Casimir energy, obtaining the Casimir entropy of the system is possible by taking the derivative in terms of the temperature as follows
\begin{equation}
    S=-\frac{\partial E}{\partial T}.
    \label{s1}
\end{equation}
Note that any null temperature contribution vanishes due to the derivative; then, we only consider terms which are temperature dependent. However, from Eq. \eqref{t15}, we will get a divergent value when applying the sums after the derivations. An alternative is getting approximations to the temperature limit, those made in the determination of the energy. Although it works, the calculus may be exhaustive and repetitive, consequently, we can apply the derivative to the energy at low- and high-temperature limits. Hence, we start with the low-temperature limit, which is expressed as
\begin{equation}
    S_{\rm low}=\pi  k_{\rm B} \ln\left(e^{-\frac{3 \pi  \beta  \hslash c}{a}}+1\right)+\frac{3 \pi ^2 \beta  \hslash c k_{\rm B} }{ a
   \left(e^{\frac{3 \pi  \beta 
   \hslash c}{a}} +1\right)^2}\left[
   \left(e^{\frac{3 \pi  \beta \hslash c}{a}}+1\right)+\frac{27 \pi ^2 \beta\bar{\alpha}  \hslash c}{4 a^3}
   e^{\frac{3 \pi  \beta  \hslash c}{a}}\right]\,,
   \label{s2}
\end{equation}
where in the limit $T\to0$, the entropy goes to zero, respecting the Nernst heat theorem as shown in Fig. \ref{entroL}. It is worth highlighting that in the regime where $\textcolor{blue}{a}\to0$ the Casimir entropy diverges as expected due to the nature of the phenomenon.

The behaviour presented in Fig. \ref{entroL} shows us an upper limit in the value of the entropy and decreases after a specific temperature. This may be explained by the loss of the quantum characteristic due to the increase in temperature. Moreover, the entropy goes to a constant value when the temperature $T\to\infty$, but it would be correct if the Casimir energy had a linear temperature dependence. This fact shows us the limitation in describing the low and high temperatures with the same expression. 

\begin{figure}[h!]
    \centering
    \includegraphics[width=0.5\linewidth]{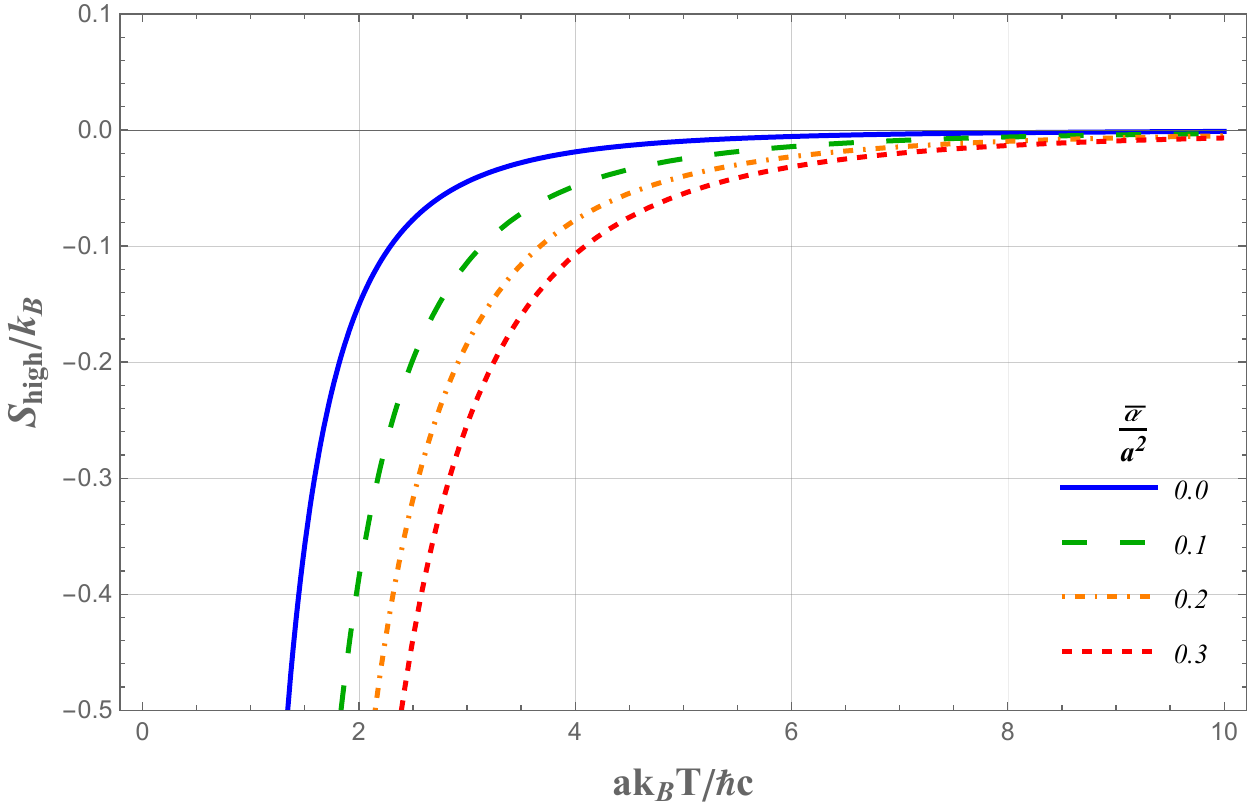}
    \caption{Plot of the dimensionless entropy $S_{\rm cas}^{\rm high}(T,a)/k_{\rm B}$ in terms of $ak_{\rm B}T/\hslash c$ with a fixed distance $a$.}
    \label{entroH}
\end{figure}

The correct approach to this limit is by using the high regime temperature, which can be straightforwardly obtained from Eq. \eqref{h8}, leading us to
\begin{equation}
   S_{\rm high}= -\frac{2 \beta ^3 (\hslash c)^3 k_{\rm B} }{a^3}\left(\frac{7 \pi ^3}{360}+\frac{31 \pi ^5 \bar{\alpha} }{1008 a^2}\right)\,,
   \label{s3}
\end{equation}
and its behaviour is shown by Fig. \ref{entroH}.

The behaviour presented by Fig. \ref{entroH} shows that the Casimir entropy vanishes as the temperature increases. This result is expected and agrees with the behaviour of the Casimir energy at high temperatures \eqref{h8}, where it converges to a constant value after reaching a critical temperature.

In summary, the entropy in Casimir systems quantifies the role of thermal fluctuations of the quantized field subject to boundaries. Its maximum corresponds to the crossover between the quantum (vacuum-dominated) and classical (thermal-dominated) regimes of the Casimir interaction. Discussing this provides a thermodynamic interpretation and clarifies that the maximum is not pathological, but a natural feature of the finite-temperature Casimir effect.


\section{SSH Model and Casimir Boundary Conditions}
\label{sec06}

To simulate Lorentz-violating fermionic dispersion in a condensed matter setting, we consider a well-established one-dimensional lattice model called the Su–Schrieffer–Heeger (SSH) model \cite{su1979solitons}. This model describes electrons hopping along a chain with alternating hopping amplitudes and captures the essential low-energy features of 1D Dirac fermions with sublattice (chiral) symmetry.

The momentum-space SSH Hamiltonian takes the form
\begin{equation}
H(k) = [t + \delta t \cos(ka)]\,\sigma_x + \delta t \sin(ka)\,\sigma_y,
\label{l1}
\end{equation}
where $t$ is the average hopping amplitude, $\delta t$ encodes the dimerization strength, $a $ is the lattice spacing, and $\sigma_{x,y}$ are Pauli matrices acting on the sublattice degrees of freedom. This Hamiltonian possesses a Dirac point at $k = 0$ (or $k = \pi/a $, depending on the sign of $t$ and $\delta t $), where the energy spectrum becomes linear and gapless in the continuum limit.

To focus on the emergence of nonlinear corrections and simplify the analysis, we adopt the limit of strong dimerization, $t \ll \delta t$. In this regime, one hopping dominates, and the chain effectively decouples into weakly connected dimers. Near the Dirac point ($ka \ll 1$), we expand Eq.~\eqref{l1} and compute the energy spectrum:
\begin{align}
\omega(k) & = \sqrt{[t^2+\delta t \cos(ka)]^2 + [\delta t \sin(ka)]^2} \notag \\
& \approx \sqrt{t^2+ (\delta t^2) a^2 k^2 \left(1 - \frac{1}{3} a^2 k^2 + \frac{2}{45} a^4 k^4 \right) + \mathcal{O}(k^6)}.
\label{l2}
\end{align}
Since the first term is suppressed by the following terms multiplied by $\delta t^2$, then the low-energy dynamics are governed by the effective Hamiltonian
\begin{equation}
H_{\text{eff}}(k) = \hbar v_F \sqrt{ k^2\left(1 - \frac{1}{3} a^2 k^2 + \frac{2}{45} a^4 k^4 \right)},
\label{l3}
\end{equation}
where $\hbar v_F = a\delta t$. This expression introduces nonlinear corrections to the linear Dirac spectrum.

For simplicity and comparison with Lorentz-violating (LIV) dispersion relations, we use a similar polynomial obtained to $\eta \equiv 0.41a$, which defines an effective nonlinearity parameter and leads us to:
\begin{equation}
\omega(k) \approx \hbar v_F \sqrt{ k^2 \left(1 - \eta^2 k^2 \right)^2 },
\label{l4}
\end{equation}
which is structurally equivalent to the LIV-inspired dispersion relation given by Eq. \eqref{f15}, i.e.:
\begin{equation}
\omega(k) \approx \hbar v_F |k| \left|1 - \eta^2 k^2\right|.
\label{l5}
\end{equation}
This effective model thus provides a lattice-based realization of nonlinear fermionic dispersion with a naturally negative $ \eta $ coefficient, consistent with compressive strain or lattice discretization effects.

To study the Casimir effect in this context, we return to the position-space form of the SSH Hamiltonian and impose open boundary conditions (OBCs) on a finite-length chain of length $L = aN$, where $ N $ is the number of sites. These boundary conditions discretize the momentum modes and allow us to compute the vacuum energy shift due to finite-size confinement, in direct analogy to the standard Casimir effect.

The quantized momenta consistent with OBCs are
\begin{equation}
k_n = \left(n + \frac{1}{2}\right)\frac{\pi}{L}, \quad n = 0, 1, 2, \dots
\label{l6}
\end{equation}
Substituting these into the nonlinear dispersion, we obtain the discrete energy levels:
\begin{equation}
\omega_n = \hbar v_F \left(n + \frac{1}{2}\right)\frac{\pi}{L} \left|1 - \eta^2 \left(n + \frac{1}{2}\right)^2 \left( \frac{\pi}{L} \right)^2 \right|.
\label{l7}
\end{equation}
The zero-point energy of the system is given by the sum over all allowed modes:
\begin{equation}
E_0(L) =- \sum_{n=0}^{\infty} \omega_n = -\hbar v_F \sum_{n=0}^{\infty} \left(n + \frac{1}{2}\right)\frac{\pi}{L} \left|1 - \eta^2 \left(n + \frac{1}{2}\right)^2 \left( \frac{\pi}{L} \right)^2 \right|.
\label{l8}
\end{equation}
Defining the dimensionless LIV parameter  $\bar{f}^2 = \eta^2 \pi^2 / L^2$, we rewrite the Casimir energy in compact form:
\begin{equation}
E_{\rm cas}(L) = -\hbar v_F \sum_{n=0}^{\infty} \sqrt{ \bar{f}^2 \left(n + \frac{1}{2}\right)^2 \left[ 1 - \bar{f}^2 \left(n + \frac{1}{2}\right)^2 \right] }=-\frac{0.41a\pi\sqrt{\delta t^2-t^2}}{24L}\left[1+\frac{1.18 \pi ^2 a^2}{40 L^2}\right].
\label{l9}
\end{equation}
This result matches the analytical structure obtained in previous studies of Casimir energy in Lorentz-violating field theories, with the LIV effects naturally emerging from the underlying lattice model. Note also that the Casimir energy scales with $\sqrt{\delta t^2-t^2}$, which determines both the Fermi velocity $v_F$ and the bulk gap. Thus, $E_{\rm cas}$ is controlled by the hopping imbalance $\delta t-t$.

\begin{figure}[h!]
    \centering
    \includegraphics[width=0.5\linewidth]{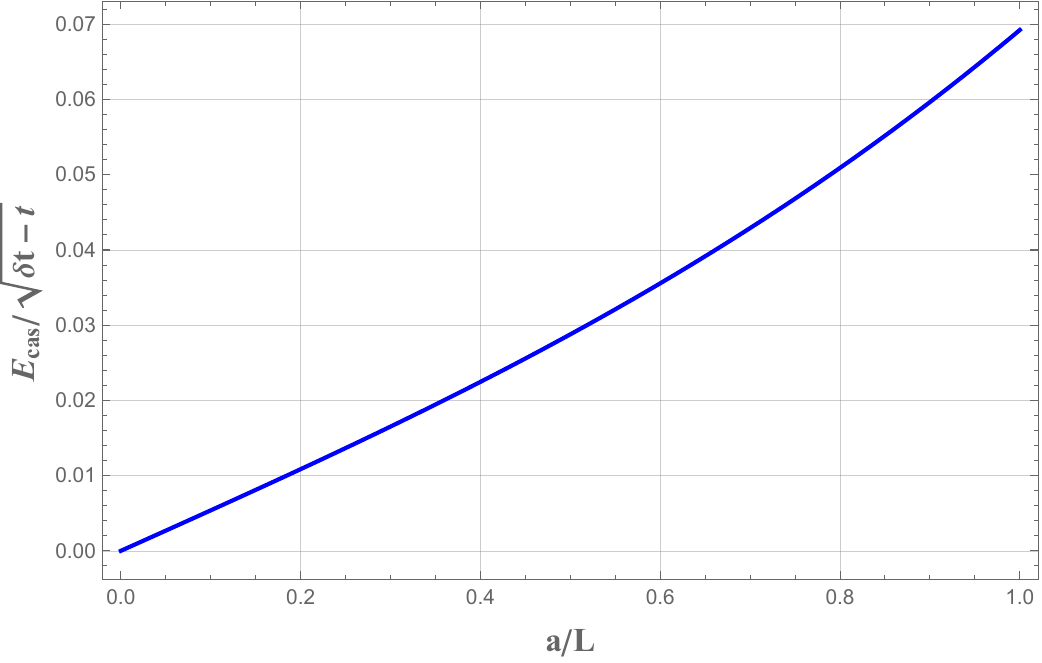}
    \caption{Plot of dimensionless Casimir energy $E_{\rm cas}/\sqrt{\delta t -t}$ in terms of a/L, highlighting the suppression of vacuum energy as the system length increases}
    \label{figlat}
\end{figure}

As shown in Fig. \ref{figlat}, the behaviour of the Casimir energy has the maximum value when the system length $L$ is small $(a/L\to1)$, and in the case of $L\to\infty$ $(N\to\infty)$ it vanishes. It occurs because in the limit of large $N$, corresponding to a macroscopic SSH chain, the quantization of momentum modes becomes increasingly dense and the influence of the boundaries on the vacuum energy diminishes. Consequently, the Casimir energy vanishes, consistent with the physical intuition that finite-size effects become negligible in the thermodynamic limit. This behaviour is consistent with the physical interpretation that the Casimir effect arises from finite-size confinement of quantum modes, which disappears in infinite systems.



\section{Conclusion}\label{sec07}

In the present work, we analyzed a fermionic system with Lorentz invariance violation (LIV) described by a CPT-even extension. From the modified dispersion relation, we considered two distinct cases: timelike and spacelike LIV, given respectively by \eqref{f12} and \eqref{f13}. Our analysis shows that only the spacelike approach is physically consistent, as it allows taking the limit $\bar{\alpha} \to 0$ smoothly, as discussed in \eqref{f15}. Using the physically viable dispersion relation \eqref{f17}, we computed the quantum vacuum fluctuations under the MIT bag boundary condition represented by Eq. \eqref{c8}. The resulting expression for the Casimir energy is exact and obtained without approximations, contrasting with results commonly found in the literature. Furthermore, the Casimir force was derived from the energy via Eq. \eqref{f1}.

In the second part of this work, the temperature was introduced using the Matsubara formalism, allowing us to derive an expression to the Casimir energy with finite temperature \eqref{t15}. However, this expression cannot satisfy the entire temperature regime since, by performing both sums, the energy diverges; consequently, two temperature regimes were adopted. The low-temperature regime, which describes the behaviour of the energy with a temperature near zero given by Eq. \eqref{l4}, is illustrated in Fig. \ref{eneL}. It is straightforward that the energy goes to zero when the temperature $T\to0$, as expected. In addition, the Casimir force is represented at low temperature by Eq. \eqref{l5} and its behaviour is shown in Fig. \ref{forcL} where it is evident that the Casimir force vanishes as $T\to0$.

The high-temperature regime is shown in Fig. \ref{enerH}, which is expressed by Eq. \eqref{h8}, and it presents a correct behaviour when $T\to\infty$ converging to a constant value. Moreover, the LIV parameter modifies the intensity of the force and increases as the value of $\bar{\alpha}$ increases. The expression of the Casimir force was found
in Eq.\eqref{h9} and its behaviour was depicted in Fig. \ref{forH}.

Another point analyzed was the entropy associated with the vacuum fluctuation. Expressions for both the low- and high-temperature regimes were derived and given by \eqref{s2} and \eqref{s3}, respectively. The behaviour at low temperature was plotted in Fig. \ref{entroL}, where the asymptotic limit to $T\to0$, respects the third law of thermodynamics, leading us to a null Casimir entropy. In the high-temperature limit, shown in Fig. \ref{entroH}, we can see that the Casimir entropy goes to zero since the energy remains constant after reaching critical values for the temperature, as described in Fig. \ref{enerH}. In both cases, the LIV parameter $\bar{\alpha}$ modifies not only the behaviour of the entropy but also its intensity. 

By analyzing the SSH model in the strong dimerization limit \eqref{l1}, we derived a nonlinear fermionic dispersion relation expressed by Eq. \eqref{l3} that closely mimics those arising in Lorentz-violating quantum field theories. Using this modified spectrum, we computed the Casimir energy of a finite chain under open boundary conditions, Eq. \eqref{l9}, and showed how it decreases with system length, vanishing in the thermodynamic limit. This result not only reinforces the lattice origin of LIV-like effects but also demonstrates the feasibility of probing modified quantum vacuum phenomena in controllable condensed matter systems. The dependence of the Casimir energy on the hopping imbalance and lattice spacing further emphasizes the tunability of these systems for simulating exotic dispersion structures.

Finally, the investigation of the Casimir effect and its thermal properties in a fermionic LIV scenario, along with the identification of a condensed matter analogue within the SSH model, was first developed in the present work.


{\acknowledgments}
We would like to thank CNPq, CAPES and CNPq/PRONEX/FAPESQ-PB (Grant nos. 165/2018 and 015/2019), for partial financial support. K.E.L.F would like to thank the Paraíba State Research Support Foundation FAPESQ  for financial support. M.A.A, F.A.B and E.P acknowledge support from CNPq (Grant nos. 306398/2021-4,
309092/2022-1, 304290/2020-3). J.R.L.S acknowledges support from CNPq (Grant no. 309494/2021-4), and FAPESQ-PB (Grant 11356/2024). A.R.Q work is supported by FAPESQ-PB. A.R.Q also acknowledges support by CNPq under process number 310533/2022-8.

\end{document}